\documentclass{elsart}

\usepackage{graphicx}
\usepackage[latin1]{inputenc}

\begin{document}

\begin{frontmatter}
\title
{Multifragmentation of very heavy nuclear systems (III):
fragment velocity correlations and event topology at freeze-out}
\author[ipno,buch]{G.~T\u{a}b\u{a}caru\thanksref{texas}},
\author[ipno]{M.~F.~Rivet\corauthref{cor}},
\corauth[cor]{Corresponding author - rivet@ipno.in2p3.fr}
\author[ipno]{B.~Borderie},
\author[buch,ganil,ipno]{M.~P\^arlog},
\author[ganil]{B.~Bouriquet\thanksref{canberra}},
\author[ganil]{A.~Chbihi},
\author[ganil]{J.D.~Frankland},
\author[ganil]{J.P.~Wieleczko},
\author[ipno]{E.~Bonnet},
\author[lpc]{R.~Bougault},
\author[CEA]{R. Dayras},
\author[ipno,ARTS]{E. Galichet},
\author[ipnl]{D. Guinet},
\author[ipnl]{P. Lautesse},
\author[ipno]{N.~Le~Neindre},
\author[lpc]{O.~Lopez},
\author[lpc]{L.~Manduci},
\author[CEA]{L.~Nalpas},
\author[krak]{P.~Paw{\l}owski},
\author[NAP]{E.~Rosato},
\author[laval]{R.~Roy},
\author[ganil]{S.~Salou},
\author[lpc]{B.~Tamain},
\author[lpc]{E.~Vient},
\author[NAP]{M.~Vigilante} and
\author[CEA]{C.~Volant} 

\collaboration{INDRA Collaboration}

\address[ipno]{Institut de Physique Nucl\'eaire, IN2P3-CNRS, F-91406 
Orsay Cedex, France}
\address[buch]{National Institute for Physics and Nuclear Engineering,
RO-76900 Bucharest-M\u{a}gurele, Romania}
\address[ganil]{GANIL, CEA et IN2P3-CNRS, B.P.~5027, F-14076 Caen Cedex,
  France}
\address[lpc]{Laboratoire de Physique Corpusculaire Caen 
(IN2P3-CNRS/ENSICAEN et Universit\'e) F-14050 Caen Cedex, France}
\address[CEA]{DAPNIA/SPhN, CEA/Saclay, F-91191 Gif sur Yvette, France}
\address[ARTS]{Conservatoire National des Arts et M\'etiers , F-75141
Paris Cedex 03, France}
\address[ipnl]{Institut de Physique Nucl\'eaire, IN2P3-CNRS et Universit\'e,
F-69622 Villeurbanne Cedex, France}
\address[krak]{H. Niewodniczanski Institute of Nuclear Physics, 31-342
Krakòw, Poland}
\address[NAP]{Dipartimento di Scienze Fisiche e Sezione INFN, Universit\`a
di Napoli ``Federico II'', I-80126 Napoli, Italy}
\address[laval]{Département de Physique, Université Laval, Québec, G1K 7P4
Canada}
\thanks[texas]{present address: Cyclotron Institute, 
Texas A\&M University, College station, Texas 77845, USA.}
\thanks[canberra]{present address: Dept. of Nuclear Physics, The Australian
National University, Canberra, ACT, 0200, Australia.}
\begin{abstract}
Kinetic energy spectra and fragment velocity
correlations, simulated by means of stochastic mean-field calculations,
are successfully confronted with experimental data for single
multifragmenting sources prepared at the same excitation energy per
nucleon in 32 AMeV $^{129}$Xe+$^{nat}$Sn and 36 AMeV $^{155}$Gd+$^{nat}$U
central collisions. Relying thus on simulations, average freeze-out times 
of 200-240~fm/$c$ are estimated
The corresponding spatial distributions of fragments are more compact for
the lighter system ($\sim$3-4V$_0$ vs $\sim$8V$_0$). 
\end{abstract}


\end{frontmatter}


\section{Introduction\label{sect1}}

A detailed investigation of multifragmentation in heavy-ion collisions 
at intermediate energies is of great interest at present time 
in connection with phase 
transition in finite nuclear 
systems~\cite{Gro97,Ric01,Beau01,Ell02,Nat02,I46-Bor02,Cho04}.
Fused systems produced in central collisions between heavy nuclei 
allow to address rather large pieces of nuclear matter, of about 
200-400 nucleons which undergo multifragmentation~\cite{I28-Fra01,I29-Fra01}.
The present paper, presenting intra-event correlations which are highly
enlightening for the origin and the features of the process,
enlarges and completes the comparison started in reference~\cite{I29-Fra01}.
 
Two nuclear reactions, at similar available energy per nucleon were 
experimentally studied:
32 AMeV $^{129}$Xe+$^{nat}$Sn and 36 AMeV $^{155}$Gd+$^{nat}$U. 
Single multifragmenting sources at the same excitation energy per nucleon
are formed in central collisions. Their properties were examined in 
detail~\cite{I29-Fra01,T16Sal97,I39-Hud03,I40-Tab03}. 
The angular and average energy distributions of all fragments ($Z \ge 5$), 
isotropic in the c.m., are compatible with those of a thermalised 
multifragmenting source. 
The two measured average fragment multiplicities are in the ratio of the 
total charges of the systems, while the charge distributions are 
identical~\cite{I12-Riv98}.
This experimental observation can be taken as a signature either of 
a large exploration of phase space or of volume instabilities.
Up to now all global reaction characteristics, except the widths
of the individual or total kinetic energy distributions of charged products, 
were equally well described by 
statistical~\cite{T16Sal97,BouHirs99,T25NLN99,Radu02} and 
dynamical~\cite{I29-Fra01} simulations. 
The two interpretations are not contradictory since one can suppose that 
the dynamics of the collisions is sufficiently chaotic to explore enough 
of the phase space,  allowing a statistical description of the fragment 
production. Higher order charge correlations generated 
in the same dynamical approach were also successfully 
confronted with the data~\cite{I31-Bor01,I40-Tab03} for the 32 AMeV 
$^{129}$Xe+$^{nat}$Sn case. Both 
calculated and experimental results suggested an enhanced production of 
equal-size fragments, compatible with a multifragmentation induced by
homogeneous density fluctuations in the liquid-gas coexistence region 
(spinodal decomposition).

We continue a step further by examining detailed properties of fragment 
kinetic energy spectra and their intra-event correlations. The validity 
of the dynamical simulation results versus the experiment will be tested 
through these observables.  Fragment velocity correlations are
supposed to bring space and time information concerning the 
multifragmentation source~\cite{Kim92} and may thus put supplementary 
constraints on models. They may allow to trace back the event topology at 
``freeze-out'', when the fragments become free 
and feel only the Coulomb repulsion.  

The method and the data are described after the experimental selection 
of events and the detailed presentation of kinetic energy properties of
fragments. Then results of dynamical simulations are compared to 
the experimental patterns. The consequences on the 
fragment spatial distribution at the end of the elapsed time 
corresponding to their separation are discussed too. 
Conclusions are finally drawn.

\section{Experimental selection and kinetic energy spectra\label{sect2}}

The data were collected, as largely described in ref.~\cite{I28-Fra01}, 
by means of the 336 multilayer detection cells of the 4$\pi$ multidetector
INDRA~\cite{I3-Pou95}. Accurate fragment identification and energy 
calibration - crucial for this type of studies - were achieved by taking 
into account pulse height defects in the silicon 
detectors~\cite{I14-Tab99} and the influence of the quenching and of 
the $\delta$-rays on the light output of the CsI(Tl) 
scintillators~\cite{I34-Par02,I33-Par02}. The energy of the detected
products is obtained with an accuracy of 4\%. In the following, results 
relative to the kinetic properties of fragments are discussed in the 
centre of mass of the reactions and thus depend on the masses (not 
measured) attributed to the fragments. In all figures, a single mass close 
to that of the $\beta$-stability valley was attributed to each fragment 
of charge Z. However recent studies on the de-excitation of hot heavy 
nuclei have shown that the cold residues have an average mass lying on 
the evaporation attractor line~\cite{Cha98}, corresponding to neutron
deficient nuclei. The two mass formulae become different
for nuclei with charges larger than 20-25; thus the chosen relation is 
only important for the heaviest fragments discussed in the present paper. 
It was verified that the average c.m. energy of these fragments changes
at most by a few MeV when changing the hypothesis on their mass, 
which do not alter the conclusions given below.

 Among the complete experimental events: total pseudo-momentum $\geq$75\% 
of the entrance channel value (Z$_{proj}$V$_{proj}$), and total detected 
charge $\ge$80 (Xe+Sn) or $\geq$120 (Gd+U), compact single sources 
were selected with the constraint of flow angle 
$\geq 60^{\circ}$~\cite{Lec94,I9-Mar97,I28-Fra01,I29-Fra01};
indeed it was shown in previous studies that while events present the
topology of emission from two sources at small flow angles, they evolve
towards a single-source configuration above $60^{\circ}$ (see figure~9 in 
ref~\cite{I28-Fra01} and figure~1 in ref~\cite{I46-Bor02}).
A minimum 
number of three fragments (Z$\geq$5) was additionally required.
The measured average fragment multiplicities are 4.3 for Xe+Sn and 6.2 
for Gd+U~\cite{I29-Fra01}. The reconstructed sources of both systems have 
an excitation energy per nucleon $\sim$ 6.5 MeV, and  average charge 
and mass (estimated following the method described in~\cite{I29-Fra01}):
$\sim$99, $\sim$236 and $\sim$150, $\sim$378: the missing charge, 
mass and energy relative to the entrance channel values are carried away 
by high energy particles (Z$\leq$2) emitted backward and forward. 

\begin{figure}[!hbtp]
\begin{center}
\includegraphics*[width=0.75\textwidth]{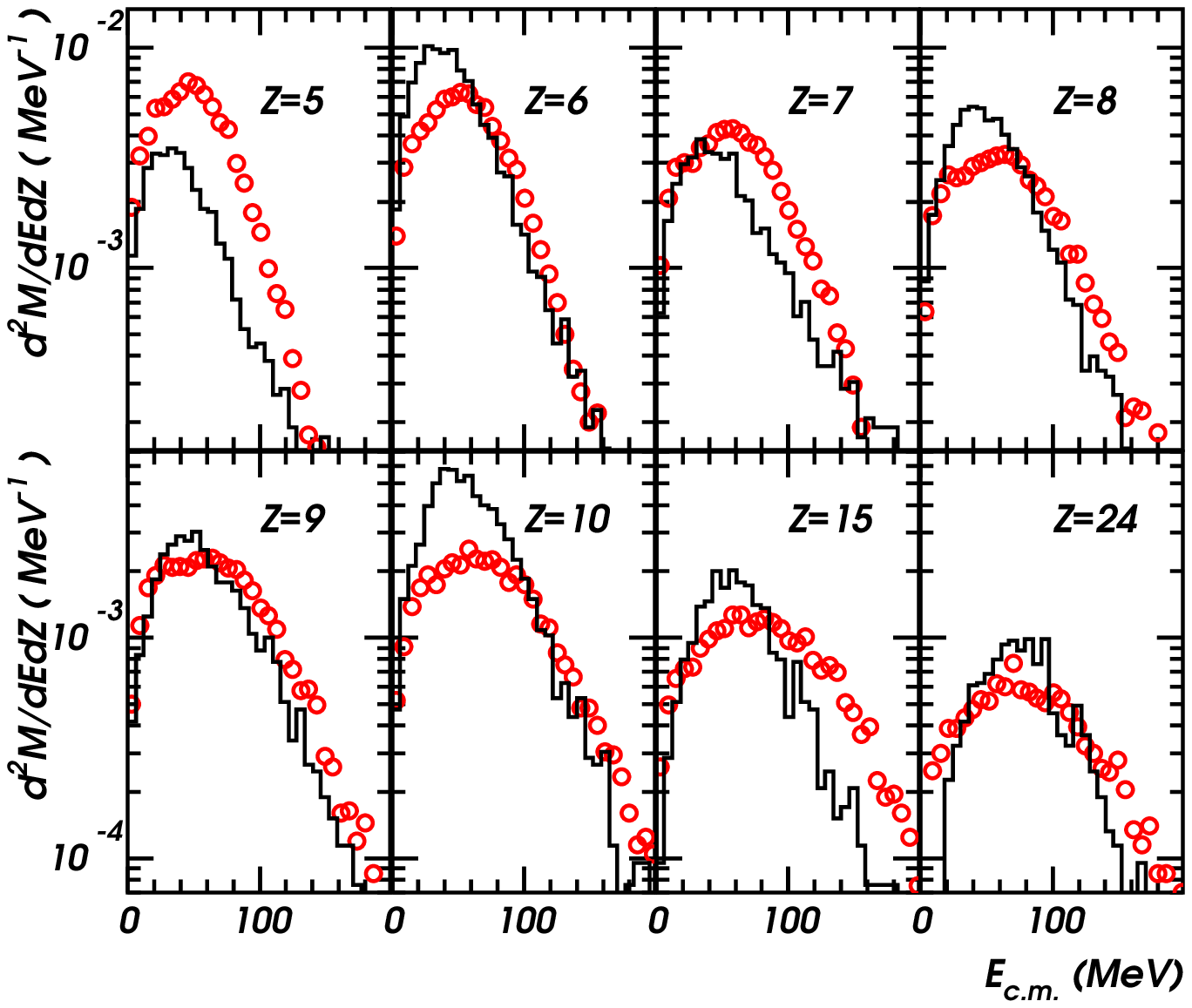} \\
\includegraphics*[width=0.75\textwidth]{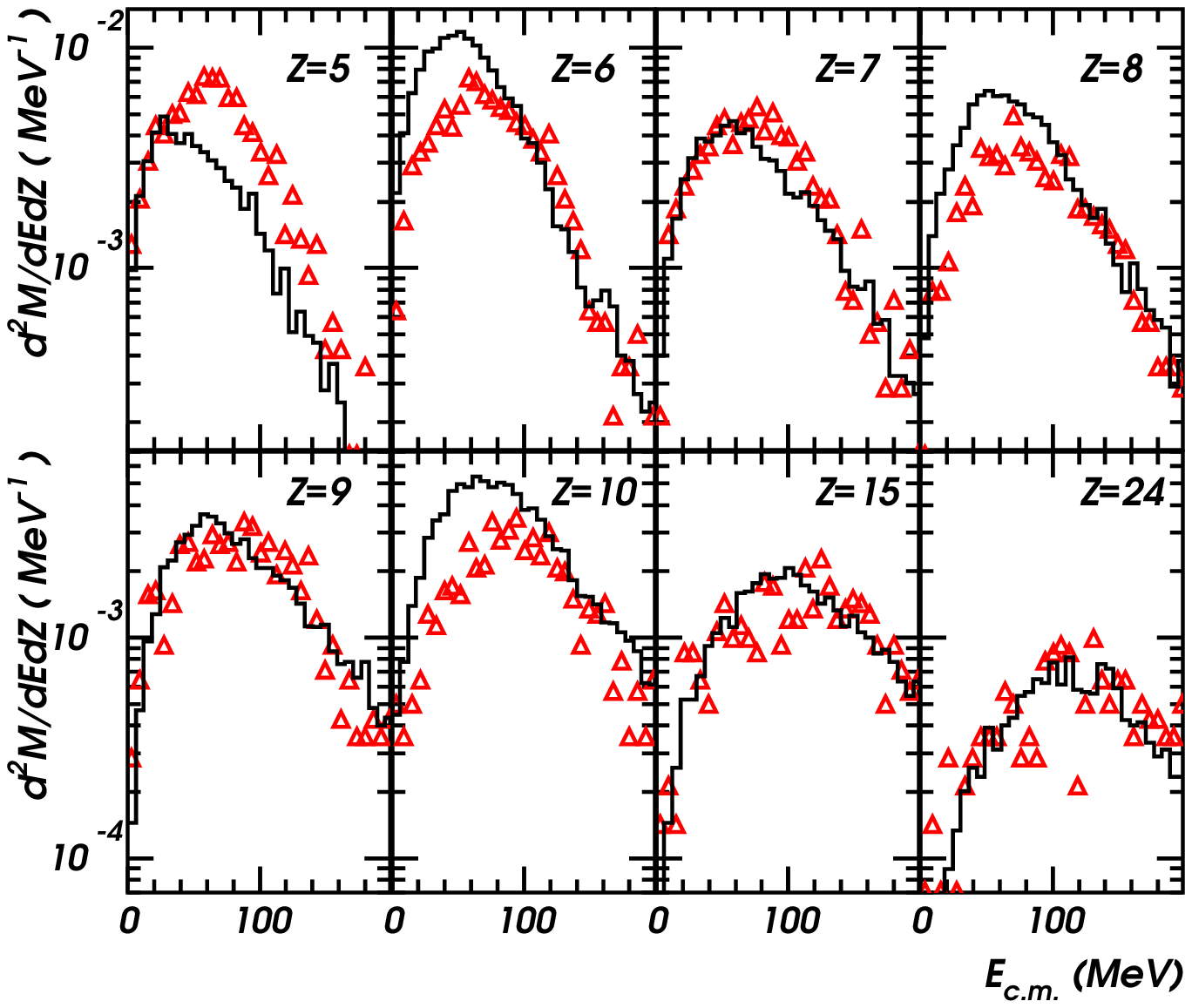}
\end{center}
\caption{Kinetic energy spectra of a few fragments for 32 AMeV 
Xe+Sn - top panel - and 36 AMeV Gd+U - bottom panel. Dynamical 
simulation results (histograms) are compared to experimental data 
(symbols).} \label{fig1}
\end{figure}
The c.m. kinetic energy spectra of various fragments emitted
in these central collisions are shown in fig.~\ref{fig1}. They present
asymmetric shapes for the lighter fragments, and tend to become more 
symmetric for Z$\geq$15, encoding information about the Coulomb 
repulsion, radial expansion and the temperature of the single source 
from which they originate. 

More information on the kinematical characteristics of the fragments
was accessed by sorting events according to the fragment multiplicity, 
M$_f$, and the rank of the fragment in the event (largest, Z$_{max}$, 
second largest, Z$_{max2}$, third largest, Z$_{max3}$, and so on).
\begin{figure}[htb]
\begin{center}
\includegraphics[width=0.8\textwidth]{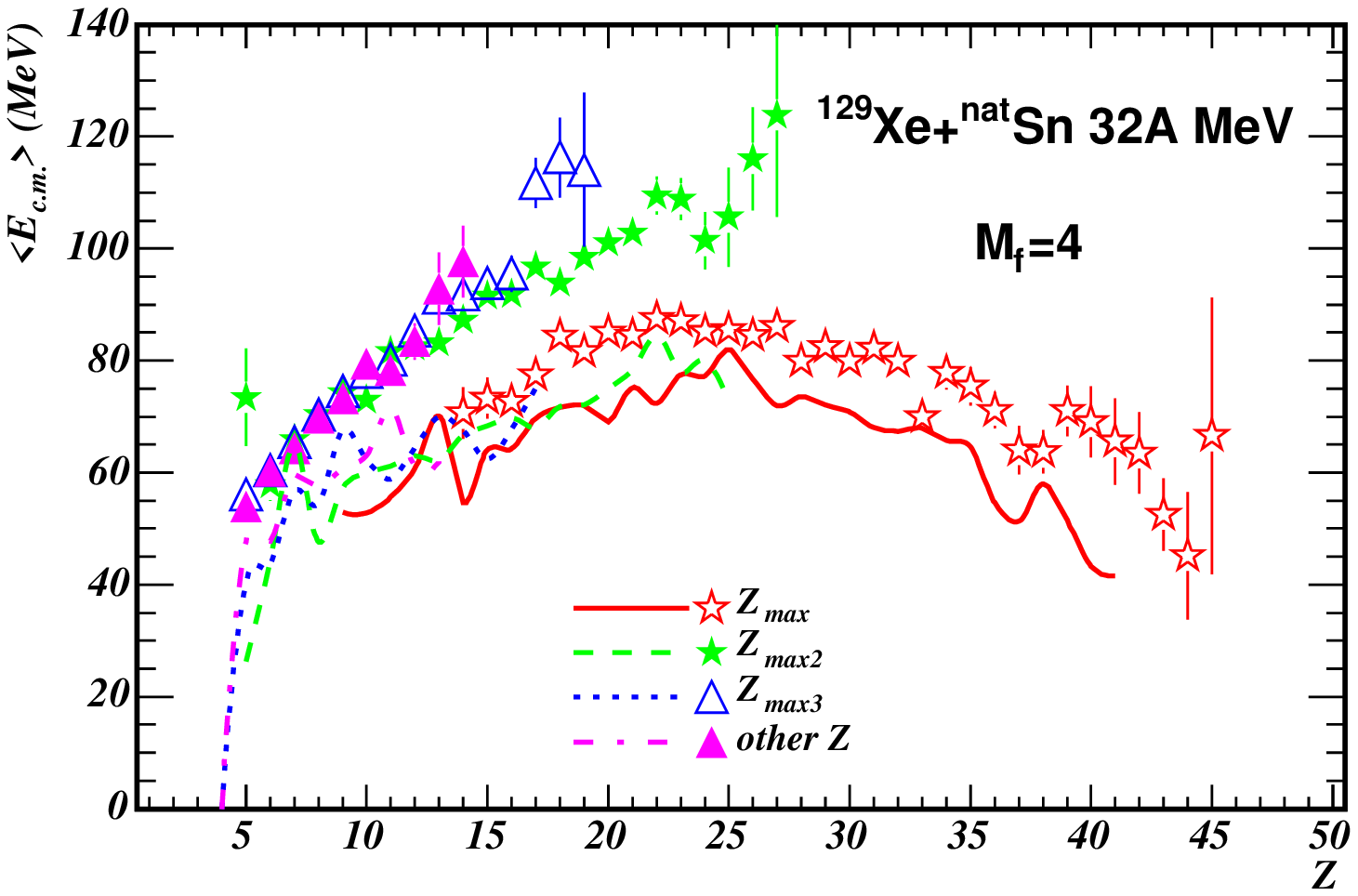}
\end{center}
\caption{Average c.m. kinetic energy of fragments in events with M$_f$=4, 
for the Xe+Sn system at 32 AMeV. Symbols show the experimental values and 
the lines the results from the BOB simulation. A minimum of 10 fragments of
given Z was required for each plotted point.} \label{EZxesn}
\end{figure}
\begin{figure}[htb]
\begin{center}
\includegraphics[width=0.8\textwidth]{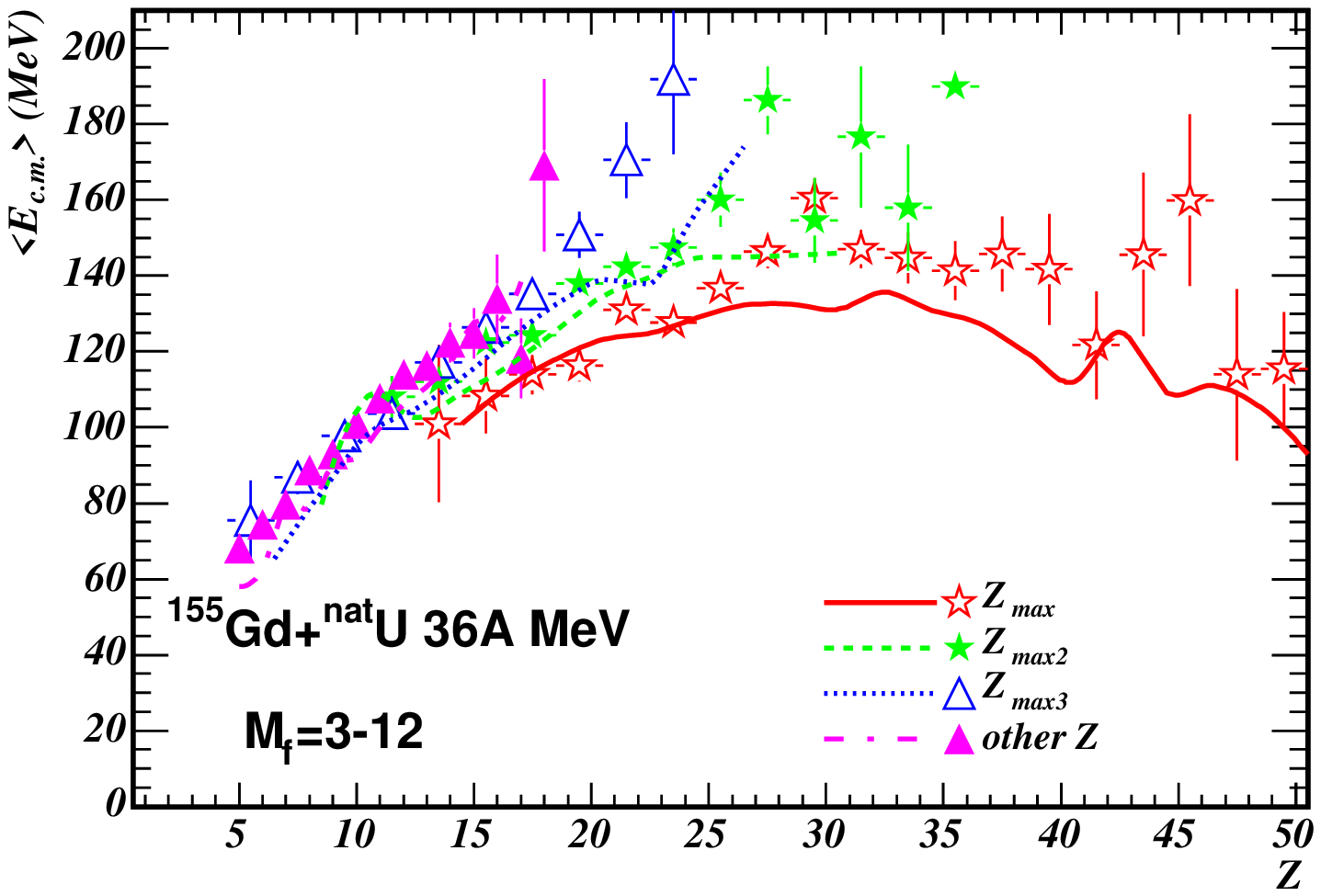}
\end{center}
\caption{Same as figure~\ref{EZxesn} for the Gd+U
system at 36 AMeV and all fragment multiplicities.} \label{EZgdu}
\end{figure}
An example of the results is shown, for the Xe+Sn system, by the symbols
in fig.~\ref{EZxesn}. Events with four fragments are displayed, but we
have observed that the experimental patterns and energy values are 
identical irrespective of the fragment multiplicity. Ref.~\cite{Col04} 
shows that, around 30AMeV, the fragment kinetic energy is
essentially from Coulomb origin; this invariance is a good 
indication that events arise from sources with very similar charge, 
independently of the fragment multiplicity. The average kinetic energy 
of fragments first increases with the fragment charge, and then 
saturates and even decreases for charges Z$\geq$20-25, independently 
of the fragment multiplicity. This evolution is a Coulomb effect. 
The most striking feature in fig.~\ref{EZxesn} is the particular role
played by the largest fragment in each partition: in the region 
Z=15-25 the heaviest fragment, Z$_{max}$, has always the lowest
average kinetic energy. Note that this behaviour was already
observed for the same system at 50 AMeV in ref.~\cite{I9-Mar97}. 
This effect may be caused by an inhomogeneity of the system
created by c.m. conservation constraints, positioning  the largest
fragment close to the centre~\cite{Rad03}; thus Coulomb  as
well as radial expansion energies would be reduced for this
largest fragment. 

\begin{figure}
\includegraphics[width=\textwidth]{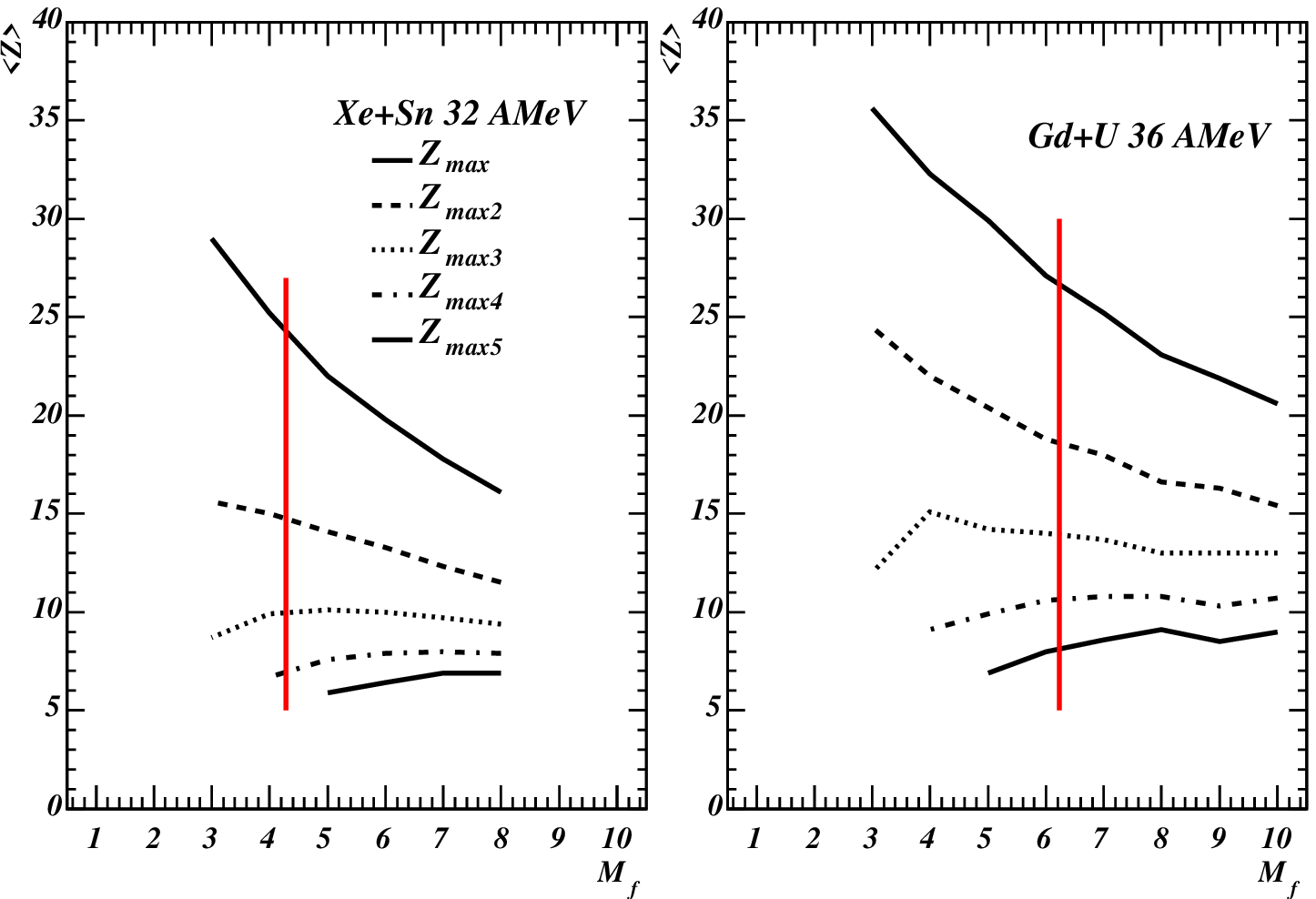}
\caption{Measured average charges of the five largest fragments of events
versus the fragment multiplicity. The vertical line represents the average
fragment multiplicity.} \label{z123}
\end{figure}
The same picture was drawn for the heavier system, Gd+U. Again one 
observes the same pattern whatever the fragment multiplicity. 
Because of the smaller number of events collected for this system, 
fig.~\ref{EZgdu} displays results without multiplicity selection. 
As expected from Coulomb effects, the average energies are larger for 
Gd+U than for Xe+Sn. Here again the average energy of the largest 
fragment is smaller than the energy of the other fragments, and shows
a maximum for Z$\sim$30-35~\cite{I29-Fra01}. However the relative 
difference between the energies of the largest fragment and those of 
the others is  smaller for Gd+U than for Xe+Sn. 

The evolution of these average energies gives first information on the 
topology of the events when nuclear interaction becomes ineffective; 
a connected observable is the average fragment charge as a function of 
their multiplicity and rank, as shown in fig.~\ref{z123}. Only Z$_{max}$ 
strongly varies with M$_f$; the fragments of rank larger than 2
keep almost constant charge values. The largest fragment again presents 
a special behaviour. The observation of figs.~\ref{EZxesn},~\ref{EZgdu},
~\ref{z123} reminds of the classification of multifragmentation 
events~\cite{Schap94} in ``sun'' (a large central fragment surrounded 
by small ones) for small M$_f$ events and ``soup'' (a random spatial 
distribution of fragments of comparable sizes) for large M$_f$ events.
However, in this last case, one would expect that the largest fragment 
looses its specificity, which is not true here as it keeps a
smaller average energy, even at large multiplicities. This above image
is thus too simple, and
one may hope to gain a deeper understanding from the velocity 
correlations presented in the next section.
Several advanced analyses took advantage of the special behaviour of 
the largest fragment, for instance universal 
fluctuations~\cite{I51-Fra05} which indicate a transition from an 
ordered to a disordered phase, and bimodality~\cite{T38Pic04,I53-Pic05} 
which suggests the occurrence of a first order phase transition in the 
studied systems.
\section{Velocity correlations\label{sect3}}

Relative velocity correlation functions between emitted products are 
tools to get information on emission time scales and on 
the disassembling source volume. Light charged particle correlation 
functions were heavily investigated in the past (see ~\cite{Bau92,Ard97} 
for reviews and~\cite{I11-Mar98,Ghe00,I39-Hud03}). A new technique, 
taking advantage of both the height and the width of the correlation 
functions was recently developed, allowing to disentangle fast and slow 
emission components~\cite{Ver02}.
Fragment correlation functions were obtained for very light 
fragments~\cite{KimY92,Bow93}, and over a much larger charge 
range~\cite{San93,San95}. In this case energy and momentum conservation
becomes important. The Coulomb interaction between the two fragments, 
and between the fragments and  the source starts to dominate; one may 
thus expect to get information on the arrangement of the fragments 
inside the source volume~\cite{Scha94}. Fragment velocity correlation 
functions are often presented as a function of a reduced velocity, 
$v_{\mathrm{red}}$, in order to increase the statistics, through mixing 
of fragments with different charges by scaling the Coulomb effects. 

Fragment velocity values are derived from their energy using estimated
masses (see sect~\ref{sect2}). The emission polar and azimutal angles 
of a fragment are chosen randomly  over those covered by the module
which detected this fragment.
For each couple $\mathrm{i,j}$ of fragments with velocities 
$\vec{v_{\mathrm i}}$ and $\vec{v_{\mathrm j}}$,
in an event of multiplicity $M_f$, ($\mathrm{i,j} = 1, ..., M_f$ 
and $\mathrm i \neq \mathrm j$), 
the reduced relative velocity is defined as~\cite{Kim92}:
\begin{equation}
\  v_{\mathrm {red}} \propto 
     \frac{|\vec{v_{\mathrm i}} - \vec{v_{\mathrm j}}|}
     {\sqrt{Z_{\mathrm{i}} + Z_{\mathrm{j}}}}
    = \frac{v_{\mathrm {rel}}}{\sqrt{Z_{\mathrm{i}} + Z_{\mathrm{j}}}}.
\label{equ1}
\end{equation}

The formalism for two particle correlation depending on one variable was 
considered for intermediate mass fragments (IMF) emitted in 
multifragmentation~\cite{Scha94,Sch94,Schap94}. 
We define the two fragment correlation function dependent on 
$v_{\mathrm red}$ as:
\begin{equation}
1 + R(v_{\mathrm {red}}) = 
\frac{\sum_{\mathrm (\vec{v_{\mathrm i}},Z_{\mathrm i}),
(\vec{v_{\mathrm j}},Z_{\mathrm j})_{\mathrm v_{\mathrm {red}}=const}} 
\prod_{\mathrm 2} \left[\left(\vec{v_{\mathrm i}},Z_{\mathrm i}\right), 
\left(\vec{v_{\mathrm j}},Z_{\mathrm j}\right)\right]}
{\sum_{\mathrm (\vec{v_{\mathrm k}},Z_{\mathrm k}),
(\vec{v_{\mathrm l}},Z_{\mathrm l})_{\mathrm v_{\mathrm {red}}=const}} 
\prod_{\mathrm 1} \left[\left(\vec{v_{\mathrm k}},Z_{\mathrm k}\right)\right]
\prod_{\mathrm 1} \left[\left(\vec{v_{\mathrm l}},Z_{\mathrm l}\right)\right]}
\label{equ2}
\end{equation}
where 
$\prod_{2}[(\vec{v_{\mathrm i}}, Z_{\mathrm i}),
(\vec{v_{\mathrm j}},Z_{\mathrm j})]$ is the probability to find two 
fragments of atomic numbers $Z_{\mathrm i},Z_{\mathrm j}$ with the 
reduced velocity $v_{\mathrm{red}}$ in one event, while the product 
$\prod_{1}[(\vec{v_{\mathrm k}},Z_{\mathrm k})] 
\prod_{1}[(\vec{v_{\mathrm l}},Z_{\mathrm l})]$
is the probability to find two fragments with the same reduced velocity, 
but in two different events. The sum at the numerator in eq. (\ref{equ2}) 
spans all couples having $v_{\mathrm{red}} = const$ in all real events 
and it accounts for correlated fragments, while the denominator accounts 
for the uncorrelated ones. The reduced velocity distribution function 
of the correlated fragments is built by taking into account all 
two-fragment combinations: $C(M_f,2) = M_f!/(2!(M_f -2)!)$ in one event 
and all the experimental selected events. For the uncorrelated case, 
we have proceeded as follows. For a given 
real event, twenty pseudo-events were generated by associating to each 
fragment another one with the same charge, randomly found in one of the 
other experimental events of the same sample. The fragment multiplicity 
and the sum of the fragment charges of the initial event are thus conserved. 
A pseudo-event is validated only if all its fragments come from different 
real events and were detected in different INDRA modules.
This procedure reduces the biases due to Coulomb effects. The ratio of 
the two distribution functions: real over pseudo-events, in 
each reduced velocity bin, leads to the correlation function defined 
in eq.~(\ref{equ2}). No normalisation factor but the factor 20 coming
from the number of uncorrelated events is used. Since only fragments from
the selected sample are used to construct the uncorrelated events, their
reduced velocity distribution essentially reflects the two-fragment
phase-space population of the detection array in the absence of any final
state interaction.

Several correlation functions are proposed here, depending on the size of 
the fragments considered in the procedure:
\begin{itemize}
\item [i)] all fragments considered ($Z_{\mathrm{i,j}} \geq 5$); 
\item [ii)] intermediate mass fragments (IMF) 
$5 \leq Z_{\mathrm{i,j}} \leq 20$;
\item [iii)] reduced velocity correlation of the heaviest fragment 
$Z_{max}$ with each of the others in the event $Z_{\mathrm i} < Z_{max}$. 
\end{itemize}

\begin{figure}[hbtp]
\begin{center}
\includegraphics*{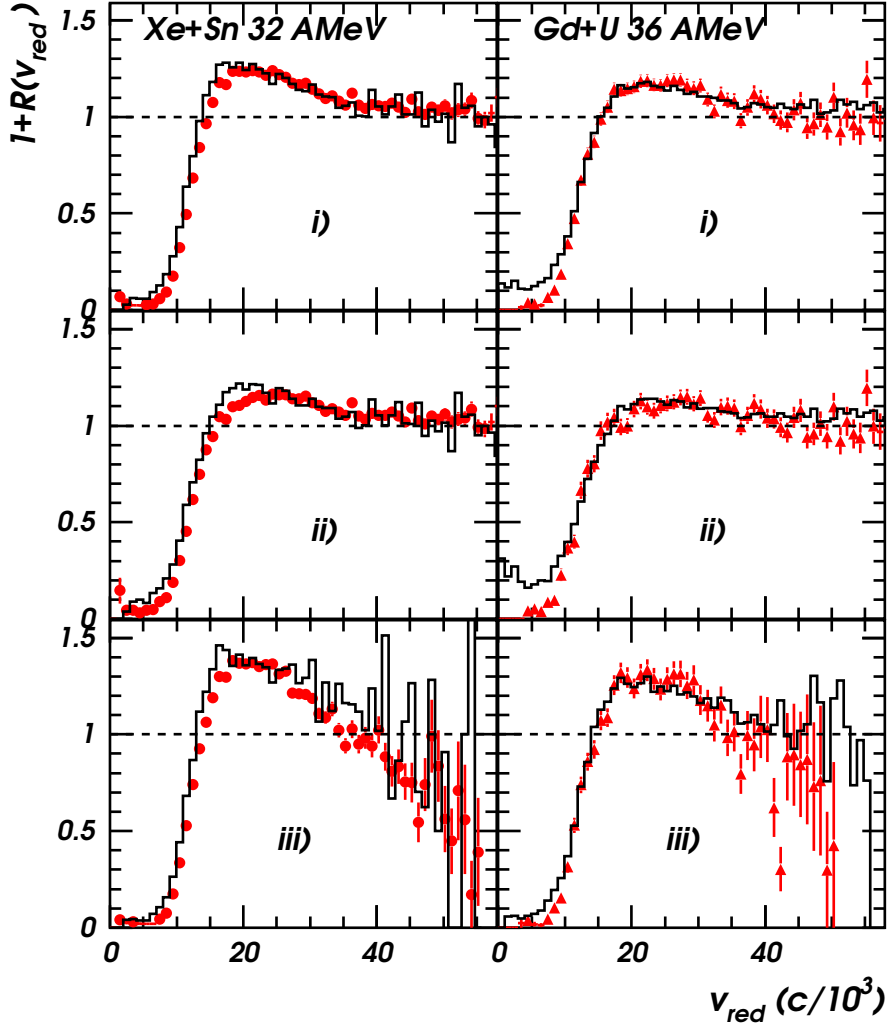}
\end{center}
\caption{Fragment-fragment  correlation functions for 32 AMeV 
Xe+Sn - left panels - and 36 AMeV Gd+U - right panels. 
Dynamical simulations (histograms) are compared to experimental data 
(solid symbols). When not visible, the statistical error bars are smaller
than the size of the symbols. Errors on the simulated functions are 
similar to the experimental errors for the Xe+Sn function of the same type.
See text for explanations of cases i)-iii).}
\label{fig5}
\end{figure}
In the absence of correlation the value of the correlation function 
should be equal to 1. At small reduced velocities momentum
conservation as well as Coulomb repulsion influence the correlation 
functions. 

The three types of experimental correlation functions are shown as
symbols in fig.~\ref{fig5}, for 32 AMeV Xe+Sn on the left side and for
36 AMeV Gd+U on the right side. It was verified with the
help of the simulation described in the next section that the
correlation functions are negligibly affected by the granularity of
the INDRA array. In the six panels one observes a dip in
the vicinity of $v_{\mathrm{red}}=0$, followed by an enhancement in the
intermediate reduced velocity region (0.015-0.035$c$). The width of the
hole, independent of the charges of the two fragments considered 
(see fig.5.14 of ref.~\cite{T16Sal97}), reflects the distances between 
fragments at freeze-out. Its depth, with a minimum value very close to 
zero, indicates that fragments are not emitted independently, as assumed 
in sequential decay models~\cite{Schap94}.
Indeed it was shown in~\cite{Tro87}, for a system similar to Xe+Sn, that 
the Coulomb interaction effect for two fragments is negligible when the 
time between their emission is long: $\sim 10^4$~fm/$c$.
The height of the bump at intermediate $v_{\mathrm{red}}$ is expected 
to be connected to the distribution 
of the nuclear matter inside the source volume at freeze-out, for 
instance, as mentioned above, ``sun events'' - evidenced by 
a very pronounced peak, or ``soup events'' for which the peak is 
diminished and the distribution is flatter~\cite{Schap94}.

The shape of the correlation function evolves, for each system,
with the size of the correlated fragments. The height of the bump at
intermediate $v_{\mathrm{red}}$ is small when all the fragments are 
considered - case i), it becomes flatter when only small fragments 
are treated - case ii). Then, it increases - case iii) 
- when the heaviest fragments are isolated in the procedure.

In each of the three cases, the width of the hole at a value  of 
the correlation function equal to 0.5 is very similar for the two systems.
The shortest intra-fragment distances seem thus to be quite independent
of the system size.
For a given system, the width of the Coulomb hole is slightly narrower
and the peak more pronounced for cases iii), testifying about a stronger 
Coulomb interaction of the heavy fragments with the others, in relation 
with their position in the source. This effect is especially visible
for the system Xe+Sn.
In fact, a more pronounced peak for this system is manifest
in all three cases, proving may be simply that, at smaller average
multiplicities, the centre of mass is closer to the heaviest fragment 
and the distances between this fragment and all the others are 
on average shorter.
While for cases i) and ii) it was verified that the correlation function
remains at 1 up to the maximum range populated by the reduced velocity
distribution ($\sim$0.09$c$), for case iii) the correlation function 
strongly decreases below 1 above $\sim$0.04$c$, particularly for Xe+Sn. 
Indeed the range of the reduced velocity distributions (both correlated and
uncorrelated) are shorter in 
that case ($\sim$0.05$c$); it was claimed 
in~\cite{Sch94} that energy and momentum conservation implies that
at very large reduced velocities, the correlation function will tend 
to zero; it is not clear why this trend  would apply only to
correlation functions of type iii). The reason of the decrease must 
probably be found in the uncorrelated distributions; it is 
indicated in~\cite{San95} that the shape of the correlation function 
is very sensitive to the background constraints when the heavier 
fragment is isolated.

Quantitative information on the source sizes and fragment emission times
will be derived from the confrontation with full dynamical simulations
of collisions in the next section.

\section{Comparisons with collision simulations\label{sect4}}

\subsection{BOB simulations\label{sect41}}
One way to explain multifragmentation in heavy ion collisions at Fermi 
energies is to correlate it with a phase transition of liquid-gas type,
due to the specific form of the nucleon-nucleon interaction.  
After a compression stage, the nuclear system enters an expansion phase, 
during which it cools down and evolves in the spinodal region (of negative 
compressibility) of the phase coexistence domain,  where 
multifragmentation occurs through the growth of density 
fluctuations~\cite{Ber83}. Stochastic mean field simulations of 
nucleus-nucleus collisions, based on the Boltzmann-Langevin equation, 
allow for the treatment of unstable systems~\cite{Ran90,Cho91,Bur91}. 
Nevertheless applications to 3D nuclear collisions are prohibited by 
computational limitations. The dynamical path through the spinodal 
region has been instead simulated by means of a Brownian 
force~\cite{Cho94,Gua97} - Brownian One Body (BOB) dynamics - grafted, 
at the instant of maximum compression, $\sim$~40~fm/$c$, on the 
one-body density evolution calculated in a Boltzmann-Nordheim-Vlasov 
(BNV) approach. The chosen self-consistent mean field potential gives 
a soft equation of state. The ingredients of the BOB simulations are 
presented in detail in~\cite{I29-Fra01}, as well as the comparison 
between average observables concerning filtered simulated events
and experimental ones.
In the simulations of head-on collisions, both systems form a single source
which expands with time and breaks-up into several fragments (as shown by 
the time evolution of the densities, figure~1 of ref~\cite{Par05}).
At 100~fm/$c$, at low density inside the spinodal zone, the systems 
are already thermalised, with a temperature of 4~MeV and a small 
radial velocity at the surface ($\sim 0.1c$), and the first 
fragments appear. An algorithm for reconstructing fragments is 
applied at intervals of 20 fm/$c$, based on minimum density cut-off 
$\rho_{min} \geq$~0.01~fm$^{-3}$. 
\begin{figure}[!hbt]
\includegraphics*[width=0.5\textwidth]{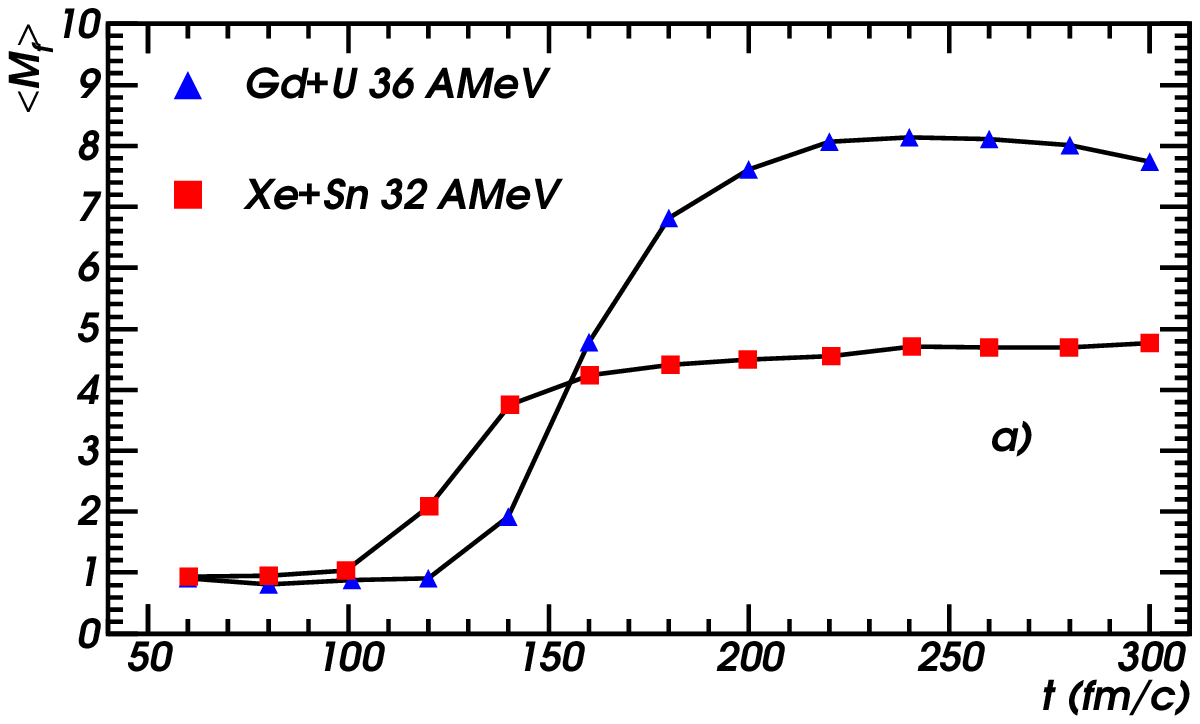}
\includegraphics*[width=0.5\textwidth]{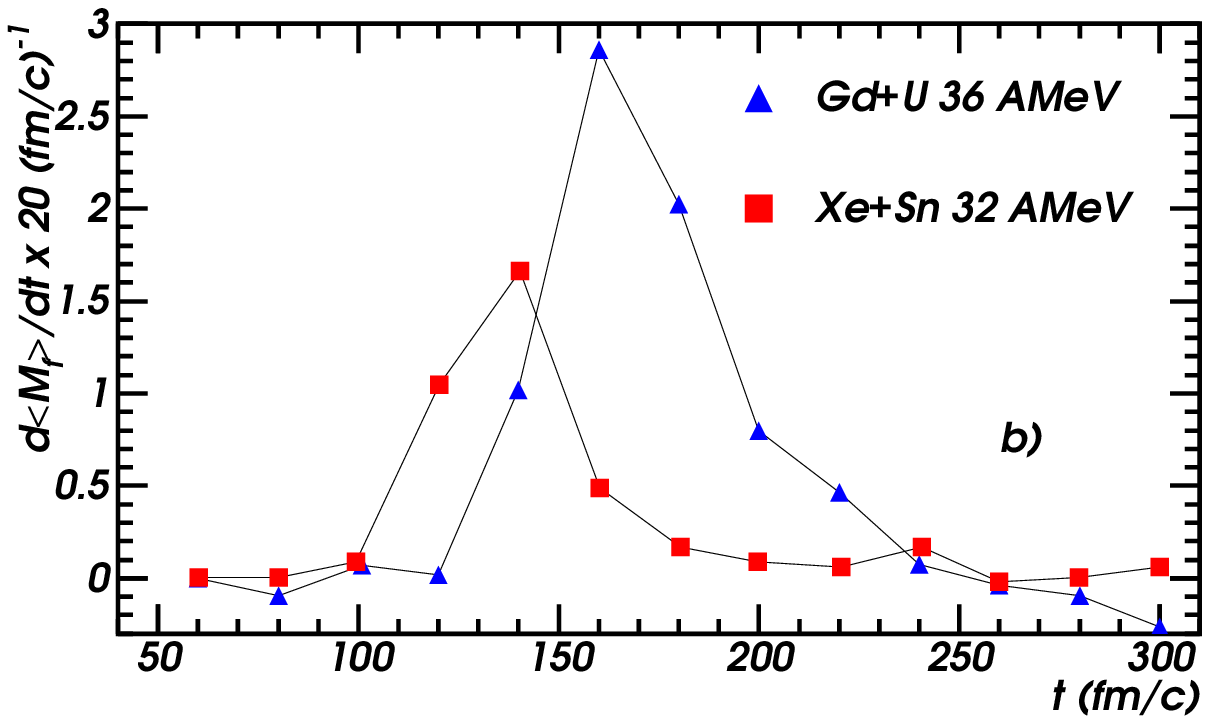}
\caption{a) Evolution with time of the average fragment 
($Z \geq 5$) multiplicity calculated for Xe+Sn and Gd+U 
collisions~\cite{I29-Fra01}; b) variation rate of the same quantity.}
\label{fig6}
\end{figure}
The calculation is stopped at the time, different for each system,
when the average fragment multiplicity becomes a constant,  as
shown in fig.~\ref{fig6}a). For a given system, this multiplicity is
independent of the value of $\rho_{min}$~\cite{I29-Fra01}.
The variation rate of the fragment multiplicity shows in 
fig.~\ref{fig6}b) that any dramatic 
change in the average number of fragments ends at $\sim$~200~fm/$c$ for 
$^{129}$Xe+$^{nat}$Sn and at $\sim$~240~fm/$c$ for $^{155}$Gd+$^{nat}$U. 
These values can be considered as reasonable average freeze-out
times~\cite{Par05}.

At the end of a BOB calculation, the fragments, still hot
($\epsilon^{*} \approx$~3~AMeV), de-excite via secondary particle 
emission, treated by means of the suitable part of the code 
SIMON~\cite{T18ADN98}. Global observables, like $M_f$, $Z$, $Z_{max}$ 
and $Z_{bound}=\sum Z_i$ distributions are very well reproduced by BOB 
calculations, while the average fragment kinetic energy is 
underestimated by about 20\% over all the Z range for Xe+Sn, and 
for Z$\geq$20  for Gd+U~\cite{I29-Fra01}. 
Part of the discrepancy might come from the semi-classical
treatment of BOB which underestimates the radial energy - by 
about 50\% - as compared to a quantal treatment~\cite{Cho04}. 

\subsection{Average energies and correlation functions\label{sect42}}

After filtering the calculated events through a software replica of 
the INDRA array, the average calculated kinetic energies of fragments 
were sorted following the fragment multiplicity and rank, as done for 
experimental data. The results are displayed by the lines in 
figs.~\ref{EZxesn} and~\ref{EZgdu}. As mentioned above, the calculated 
energies underestimate the experimental values for Xe+Sn while they 
agree with data for the lighter elements in Gd+U. 
Besides a drawback in the calculation mentioned above, this may sign 
a more compact shape of the experimental configuration 
with respect to the calculated one for Xe+Sn. In agreement with 
experiment the average energy for a given charge is independent
of the fragment multiplicity (not shown) and the average energy of the 
largest fragment decreases when its charge becomes larger than 
$\sim$25 (Xe+Sn) or $\sim$30-35 (Gd+U). For the heavier system, the 
calculated energies show the same hierarchy with the fragment rank, 
namely the energy of a given charge is smaller when the fragment is 
the heavier of the partition. Conversely, the calculated energies are 
independent of the fragment rank for Xe+Sn. This may indicate that the 
topology of the simulated events reflects well the real one for Gd+U 
while it is slightly different for Xe+Sn. A different topology also
influences the calculated values of the fragment energy:
as stated in~\cite{Cho04}
Coulomb acceleration is more effective for a uniform distribution than 
for a hollow configuration as that obtained in BOB~\cite{Par05}. 
Thus both the larger absolute values of the fragment energies, and 
the smaller energy of the largest fragment may indicate that, for Xe+Sn, 
the experimental freeze-out configuration is more compact, 
more uniformly filled, 
with the largest fragment closer to the centre, than the calculated 
configuration.

We have also confronted the simulated kinetic energy spectra of the 
final fragments with the experimental ones. Fig.~\ref{fig1} shows that, 
for Xe+Sn, the calculated spectra are narrower than the experimental ones,
and more so for higher charges.  For the heavier system  the agreement 
between calculated and measured spectra is better:
both the average values and the widths of the spectra are reproduced 
to within a few percents for odd charges and about 10\% for even 
charges, while the differences on these quantities lie between 10 and 
20\% for Xe+Sn. A calculated odd-even effect can be noted: not 
only is the production of even charge fragment favoured, but both the 
average values and the widths of their spectra are farther from the
experimental values than those of the odd-charge fragments. 
 Note that for both systems fragments with charge Z=5 are 
strongly underestimated in the calculation.
While the odd-even effect has to be attributed to the de-excitation 
process, as no such effect is seen in the primary 
distributions, the disagreement between calculation and experiment,
besides the above mentioned differences in topology, may originate
from \\
i) the dropping-out of the light charged particles in the last step of 
the calculation: only fragments are input to the SIMON code, the 
particles already free (27\% of the initial charge for Xe+Sn, 
22\% for Gd+U) do not participate to the Coulomb propagation. 
This contributes to the underestimation of the energy.\\
ii) the spatial distribution of these primary particles: placed
mainly in the centre of the source instead of being uniformly 
distributed, they would increase the Coulomb effect. \\
To summarize, these dynamical calculations fit rather well the 
individual experimental energy spectra - a difficult task, 
scarcely reported up to now for other models.

 The method of reduced velocity correlations, as described in the 
previous section, is applied in the following to fragments from BOB  
simulated events. A good agreement between the simulated 
(histograms) and experimental (symbols) correlation functions is 
observed in fig.~\ref{fig5}. The depth and the width of the Coulomb 
hole are  well reproduced by the present dynamical simulations for the 
different correlation functions. Note that the decrease below 1 of 
the correlation functions of type iii) is present 
also in the simulations. For Gd+U, the agreement between the
experimental and calculated correlation functions is excellent; the 
small filling-up at low reduced velocity is introduced by the 
filtering step. For Xe+Sn, the three calculated functions are slightly 
narrower than the experimental ones and the peak is displaced towards 
smaller reduced velocities. This discrepancy confirms 
the difference between the experimental and calculated 
event topology already mentioned.
However it should be stressed that, for both systems, 
the correlation functions generated 
by the BOB simulations account better for the experimental ones 
than statistical models like SMM or MMM~\cite{T16Sal97,TabBol00,Radu02}.

\subsection{Freeze-out configuration: volume estimate and
topology\label{section43}}

The agreement between calculated and experimental kinetic properties,
especially for the heavier system, encouraged us to give signification 
to the positions of the fragments just after their formation. 
\begin{figure}[!hbt]
\begin{center}
\includegraphics*[width=0.8\textwidth]{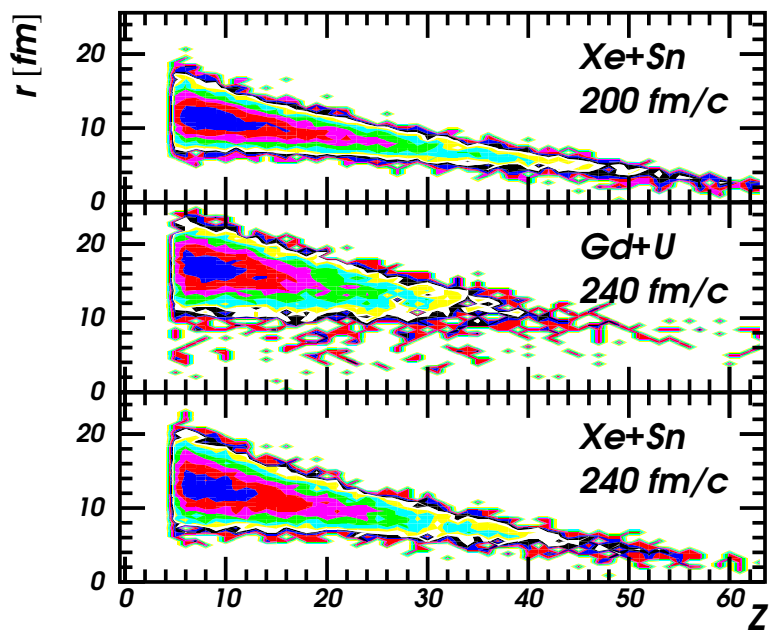}
\end{center}
\caption{Fragment positions as a function of the atomic number 
for: 32~AMeV $^{129}$Xe+$^{119}$Sn at 200~fm/$c$ from the beginning of 
BOB calculation - upper panel, 240~fm/$c$ - lower panel and 36~AMeV 
$^{155}$Gd+$^{238}$U at 240~fm/$c$ - middle panel. The contour scale is
logarithmic.}
\label{fig7}
\end{figure}
We analysed the calculated events at the moment when their multiplicity
variation rate vanishes: at 200~fm/$c$ from the beginning of the 
calculation for Xe+Sn and at 240~fm/$c$ for Gd+U (see fig. \ref{fig6}b). 
At the moment of the fragment separation, the values of their distances 
$r_{i}$ from the source c.m. or of their relative distances 
$|\vec{r_{i}} - \vec{r_{j}}|$ give information 
concerning the topology of events and associated freeze-out volumes.

\begin{figure}[hbt]
\begin{center}
\includegraphics*[width=0.8\textwidth]{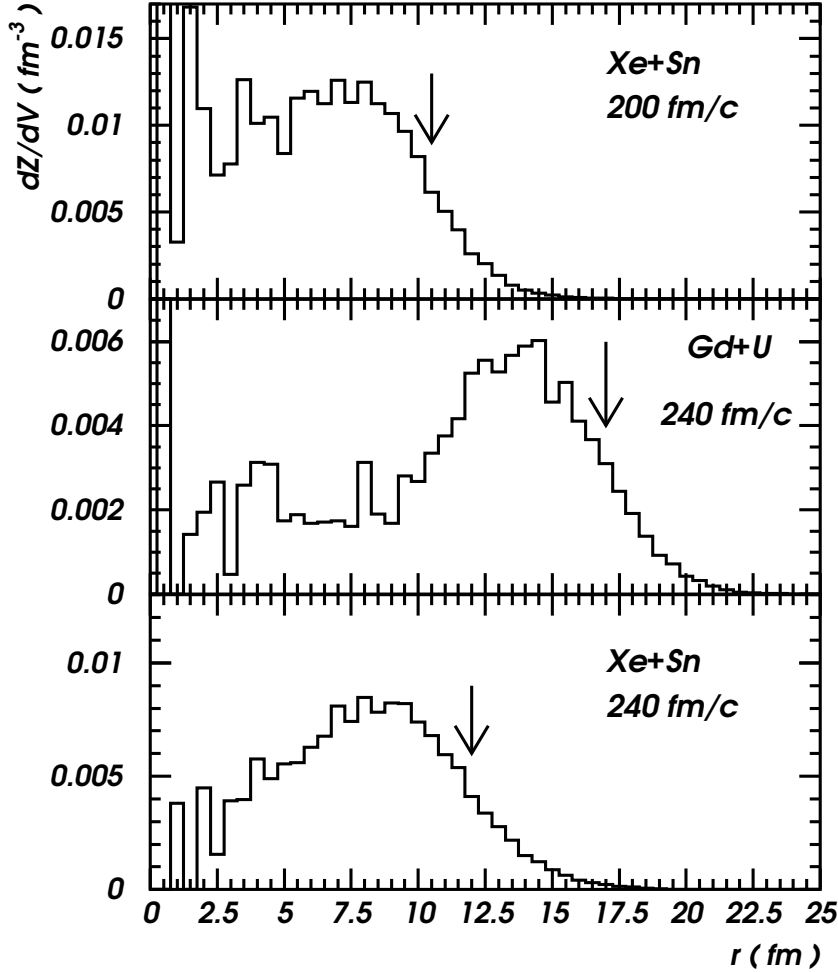}
\end{center}
\caption{Radial charge distributions.
Upper panel: 32 AMeV $^{129}$Xe+$^{119}$Sn at 200 fm/$c$ from the 
beginning of the BOB calculation; middle panel:
36 AMeV $^{155}$Gd+$^{238}$U at 240 fm/$c$; lower panel: as in the upper
panel but at  240 fm/$c$. The arrows indicate the radii used to estimate the
volumes reported in Table \ref{table1} (see text for explanations).}  
\label{fig8}
\end{figure}
The position radii $r$ of the fragments relative to the c.m
- fig.~\ref{fig7} - are shorter in the Xe+Sn reaction at 200~fm/$c$ 
(upper panel) than in Gd+U at 240~fm/$c$ (middle panel), 
which may be attributed to the shorter propagation time on one hand, 
and to the Coulomb interaction ($E_{coul}$) and the radial
expansion ($\epsilon_{rad}$), both inferior for the smallest 
source~\cite{I29-Fra01}.
The effect of the time difference is shown on the lower
panel of the figure.
The ridge lines of the plots for the two systems have the same 
slope~\cite{T28Tab00}, $r$ decreases when $Z$ increases. 
These observations explain the behaviour of the fragment-fragment 
correlation functions shown in fig.~\ref{fig5}: the heavier fragments 
take more central positions, at least at the end of the fragment 
separation time. In case of Xe+Sn system the heavier fragments are 
moreover very close to the centre of mass (2-3 fm), while they are 
farther away for Gd+U. Correlatively the peak in the correlation 
functions increases when the heavier fragments are involved and is 
more pronounced for the lighter system. 

\begin{table}[!hbt]
\caption{Some characteristics of the 32~AMeV Xe+Sn and 36~AMeV 
Gd+U single sources - calculated by means of BOB dynamical model 
\cite{I29-Fra01} - in the spinodal region (100~fm/$c$), columns 2 and 3,
and at the moment when the mean multiplicity of fragments, $\bar{M_f}$ ,
saturates. In the last column, the ratio of the average volume of a sphere 
englobing nearly all the fragment centres and the source volume 
at normal nuclear density.} 
\vspace*{0.2cm}
\begin{center}
\begin{tabular}{c|cc|cc|c}
\hline
\hline
system & $A_{tot}$ & $Z_{tot}$  & t (fm/$c$) & $\bar{M_f}$  & $ V/V_0 $ \\
\hline
$^{129}$Xe+$^{119}$Sn  & 238 & 100 & 200 & 5.1  & 2.8 \\
\hline
$^{155}$Gd+$^{238}$U  & 360 & 142 & 240 & 8.1  & 7.9 \\
\hline
$^{129}$Xe+$^{119}$Sn  & 238 & 100 & 240 & 5.1  & 4.2 \\
\hline
\hline
\end{tabular}
\end{center}
\label{table1}
\end{table} 
\begin{figure}[hbt]
\begin{center}
\includegraphics*[width=0.8\textwidth]{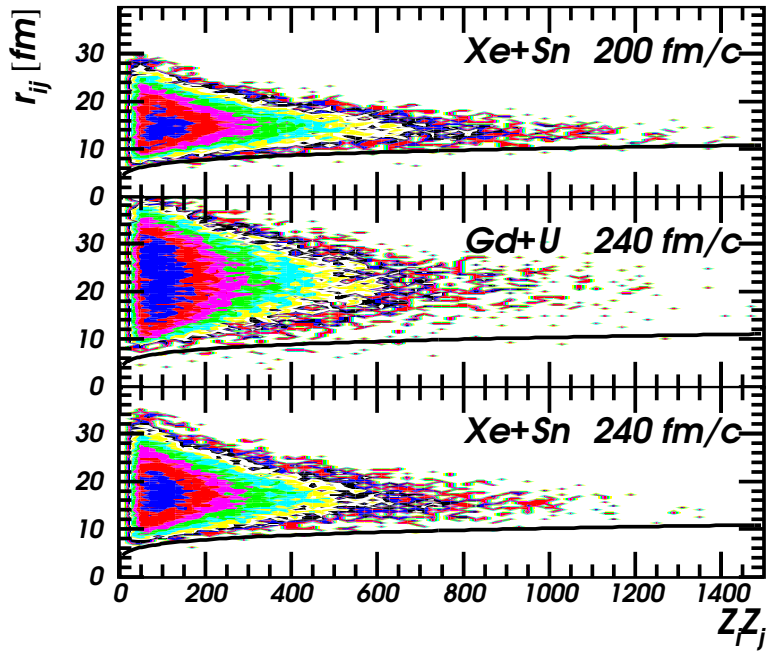}
\end{center}
\caption{Fragment relative distances vs the product of the atomic numbers 
of the fragments considered in each couple, for: 32~AMeV
$^{129}$Xe+$^{119}$Sn at 200 fm/$c$ from the beginning of BOB 
calculation - upper panel, 240 fm/$c$ - lower panel, and 36~AMeV 
$^{155}$Gd+$^{238}$U at 240~fm/$c$ - middle panel. The lines show the
minimum distances between two neighbour fragments. The contour scale is
logarithmic.}
\label{fig9}
\end{figure}
The local charge concentration per unit volume, d$Z$/d$V$, plotted 
in fig.~\ref{fig8} 
as a function of the distance $r$ from the c.m. for both systems, 
summarizes well what was discussed all along the paper: the 
configuration for Xe+Sn is more compact (and one derives from the 
experiment a configuration even more uniformly filled) than the one 
observed for Gd+U. This is also well illustrated by the relative 
distances between two fragment centres 
$r_{ij} = |\vec{r_{i}} - \vec{r_{j}}|$, shown in fig.~\ref{fig9}. The
shortest distances concern the closest neighbours, the longest ones 
are representative of the size of the sources. Relative distances 
are shorter for the reaction Xe+Sn at 200~fm/$c$ and also at 240~fm/$c$ 
- fig.~\ref{fig9} (upper and lower panels) than for Gd+U at 
240~fm/$c$ (middle panel). 
The profile of each distribution is horizontal, showing that the average 
distance between fragment centres is the same, irrespective of their size.
The minimum distance ($\sim$ 8~fm for the smallest neighbour fragments) 
is the same in all cases. The points corresponding to big fragments 
($Z_i Z_j > 700$) are close to the limit line (calculated for two equal
touching spheres at normal nuclear density) for the lighter system. 
It is not the case for the heavier system (middle panel).

Moreover fig.~\ref{fig8} can also provide a rough estimate of the
freeze-out volume. d$Z$ is the infinitesimal number of charges in the
volume element d$V = 4 \pi r^2 dr$. The upper panel in fig.~\ref{fig8}
corresponds to Xe+Sn single source at 200~fm/$c$. The distribution has
a nearly Gaussian shape, of mean $\bar{r}$ and full width at half
maximum FWHM. A sphere of radius $\bar{r} + FWHM/2$, indicated by the
arrow in the figure, englobes most of the fragment centres and has a
volume of $\sim$~2.8~$V_0$; the middle panel concerns the Gd+U single
source at 240~fm/$c$: the volume of the sphere of radius
$\bar{r} + FWHM/2$ is $\sim$~7.9~$V_0$. Volumes at normal density
$V_0 = (1.2)^3 A_{tot}$ fm$^{-3}$ are calculated for the masses of the
sources given in table~\ref{table1}. The large difference between the
volumes of the  two systems, which may look surprising, is coherent with
the influence of the system size on the Coulomb and radial
expansion~\cite{Par05}, as already suggested by the average kinetic
energies of the fragments (figs.~\ref{EZxesn} and~\ref{EZgdu}).
It might even be more pronounced, taking into account the
underestimation of the compactness of the Xe+Sn system visible in the
small discrepancy between the calculated and experimental correlation
functions.
The difference in the volume of the Xe+Sn source within 40~fm/$c$
around the average freeze-out instant (figure~\ref{fig8} and
table~\ref{table1}) gives an estimate of the uncertainty on the
determination of this volume. Note that the volume obtained at 
240~fm/$c$ for Xe+Sn compares well with that estimated by a simpler
simulation using the measured charge partitions~\cite{Ixx-Pia05}.

\section{Conclusions\label{sect5}}
Energy spectra of different fragments, issued from central collisions 
of 32~AMeV $^{129}$Xe projectile on a $^{nat}$Sn target and 36~AMeV 
$^{155}$Gd projectile on a $^{nat}$U target were presented. A detailed 
study of the evolution of the average energy versus fragment charge, 
and of the average sizes of the ordered fragments versus fragment 
multiplicity evidences the particular role of the largest fragment 
in each event. 
Experimental relative velocity correlation function of different 
types were built. Similarities between the two systems were inferred 
for the intra-fragment distances, while the position of the larger 
fragments relative to the centre of mass
seemed to depend on the size of the multifragmenting source.

All experimental results were satisfactorily well described 
by a stochastic mean-field model calculation, proving the capability 
of this dynamical approach to reproduce in detail the experimental 
data and not only global, average observables. The observed differences 
noted for the Xe+Sn system may reveal that the topology of the events 
at freeze-out is not identical to the experimental one: fragmentation 
products seem to be more uniformly  distributed in space in the 
experimental configuration. Conversely the agreement between simulation 
and experiment is excellent for Gd+U, 
allowing to state that the simulation provides plausible reaction scenario
and freeze-out topology.
In  the BOB image, involving the spinodal decomposition of the source 
(about 100~fm/$c$ after the beginning of the collision), the 
freeze-out instant may be seen as the end of a time interval in which 
all the fragments are well separated, i.e. that moment where the last 
two closest fragments move away at $\sim$2~fm. This fact is on average
accomplished at 200~fm/$c$ from the beginning of the simulations for 
the lighter system and at 240~fm/$c$ for the heavier one. 
At the same moment: 240~fm/$c$, when the fragment multiplicity is 
definitively stable in both reactions, the radial fragment distribution 
reflects the influence of the system size on the Coulomb and radial 
expansion. The freeze-out volumes estimated at this moment
are several times bigger than the volume $V_0$ of the single source 
calculated at normal nuclear density.
The higher compactness of the lighter system was quantitatively
confirmed, as well as the more central position of the
larger fragments.



\end{document}